# Beyond touch-based human-machine interface: Control your machines in natural language by utilizing large language models and OPC UA

Bernd Hofmann[a]*, Niklas Piechulek[a], Sven Kreitlein[a], Jörg Franke[a], Patrick Bründl[a]

[a] *Institute for Factory Automation and Production Systems (FAPS) Friedrich-Alexander-Universität Erlangen-Nürnberg, Germany*

* *E-mail address:* bernd.hofmann@faps.fau.de; ORDID: https://orcid.org/0009-0009-0666-6149

**Abstract**

This paper proposes an agent-based approach toward a more natural interface between humans and machines. Large language models equipped with tools and the communication standard OPC UA are utilized to control machines in natural language. Instead of touch interaction, which is currently the state-of-the-art medium for interaction in operations, the proposed approach enables operators to talk or text with machines. This allows commands such as "Please decrease the temperature by 20 % in machine 1 and start the cleaning operation in machine 2." The large language model receives the user input and selects one of three predefined tools that connect to an OPC UA server and either change or read the value of a node. Afterwards, the result of the tool execution is passed back to the language model, which then provides a final response to the user. The approach is universally designed and can therefore be applied to any machine that supports the OPC UA standard. The large language model is neither fine-tuned nor requires training data, only the relevant machine credentials and a parameter dictionary are included within the system prompt. The tool-calling ability and their design is evaluated on a demonstrator setup with a Siemens S7-1500 programmable logic controller with four machine parameters. Fifty synthetically generated commands on five different models were tested and the results demonstrate high success rate, with proprietary GPT-5 models achieving accuracies between 96.0 % and 98.0 %, and open-weight models reaching up to 90.0 %. Afterwards the approach was transferred to a deployed spay-coating machine. The proposed concept is supposed to contribute in advancing natural interaction in industrial human-machine interfaces.

## 1. Introduction

Automated machines are omnipresent in modern manufacturing environments, and human operators increasingly act as supervisors of networked cognitive systems [1]. While machines perform repetitive tasks, humans intervene whenever knowledge-based interaction or decision-making beyond predefined rule sets are required [1]. Especially in complex manufacturing contexts, the human-machine interface (HMI) has become increasingly important, with modern systems demanding fast and natural interaction [2,3]. With the technological advances in deep learning (DL) and large language models (LLM), HMIs are evolving from traditional graphical user interfaces (GUI) into natural user interfaces (NUI). NUIs enable operators to interact with computer systems through natural language or kinesics [2]. While such interaction is already state-of-the-art (SOTA) in consumer devices like smartphones, industrial applications face unique challenges: machines are expensive, potentially dangerous, and often operate in noisy environments [4]. Consequently, additional theoretical frameworks and empirical investigations are necessary to enable the broader deployment of these technologies in factories.

This paper proposes an agent-based approach for modifying machine parameters through natural language, potentially replacing or supplementing SOTA touch interaction in industrial HMIs. Figure 1 illustrates the overall concept schematically. The approach utilizes LLMs equipped with tools to convert semantic operator inputs into executable PLC (programmable logic controller) actions through an OPC UA (Open Platform Communications Unified Architecture) [5] connection. The tool-calling accuracy and their design is evaluated on a demonstrator setup with three proprietary and two open-weight LLMs across 50 synthetically generated operator commands. Furthermore, the approach is transferred to a real-world and deployed spray-coating machine.

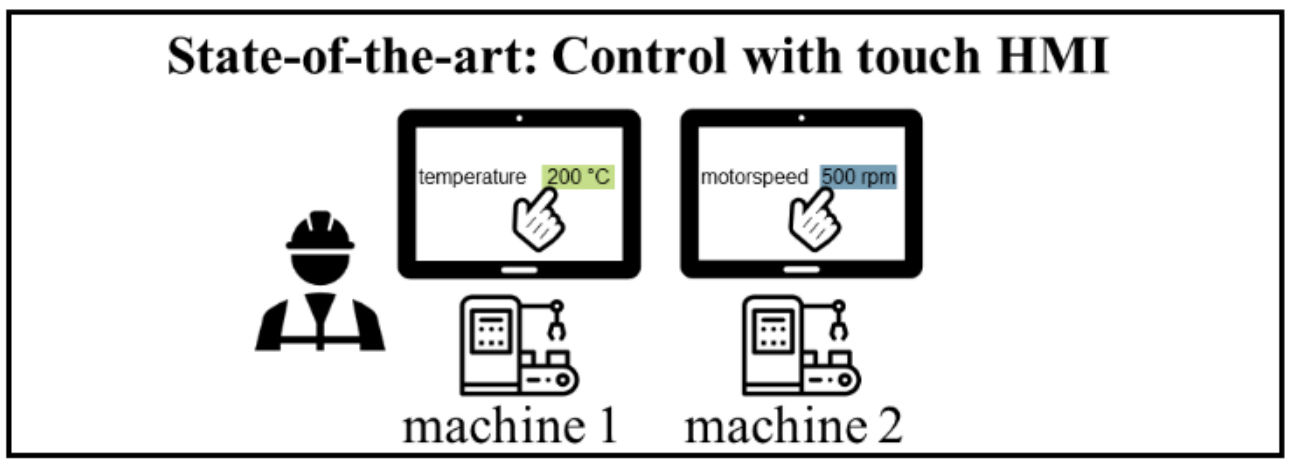

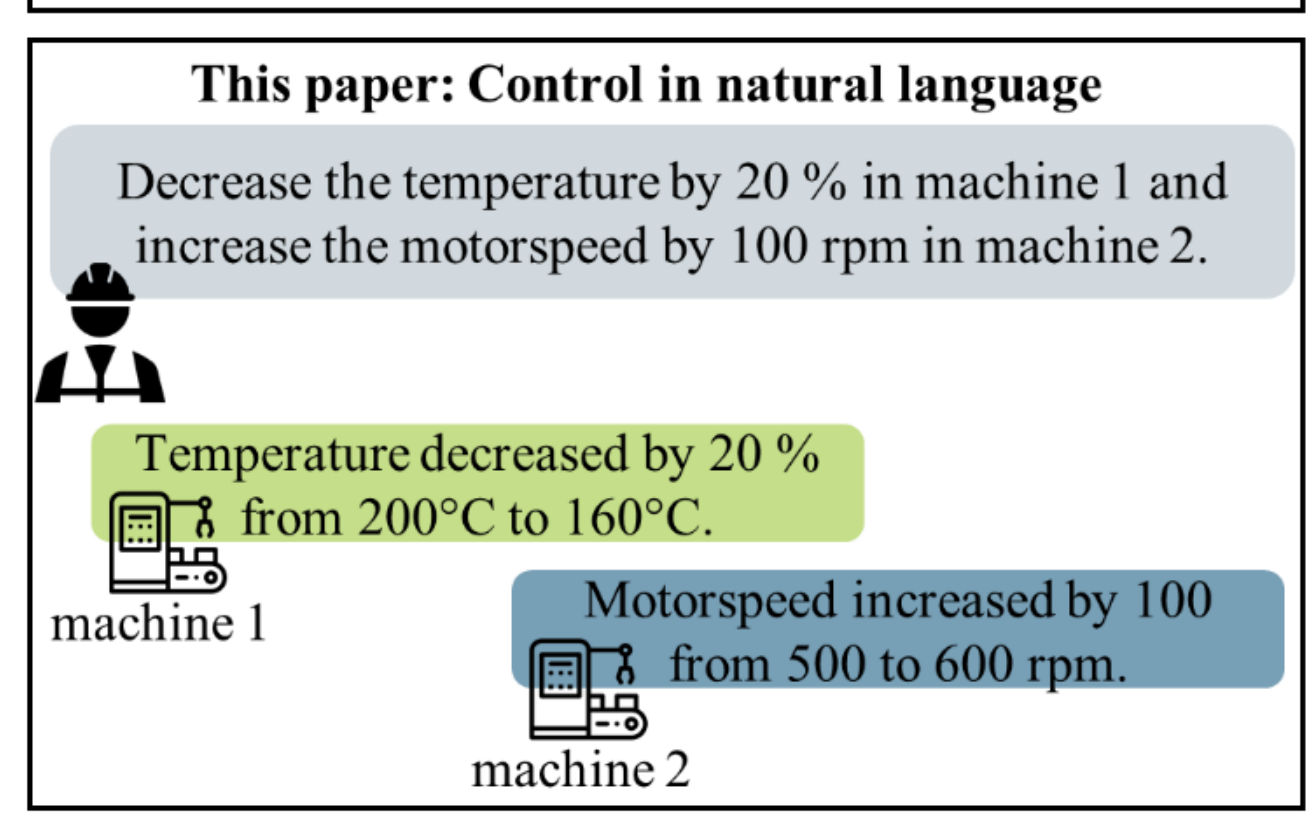

**Figure 1.** Comparison of state-of-the-art machine control using touch-based HMIs (top) and the proposed natural language-based approach (bottom).

## 2. Background

The rapid improvements in information technology have enabled humans to interact with computers in increasingly natural ways, evolving from command line interfaces (CLI) over mouse and touch-based input through GUIs to speech and gesture interaction with NUIs [6]. SOTA industrial HMIs typically rely on touch interaction and deep menu level navigation to control a machine. Such complexity can be time-consuming in daily operations and may necessitate additional operator training. NUIs aim to reduce navigation complexity, improve interaction efficiency and enable a more conversational interface between a human and the computer system. Studies across multiple domains, including industrial environments, report reduced task execution times, improved accuracy, and higher user satisfaction through natural language interaction [4,7,8]. To interact with a machine in an industrial context, the system is required to convert semantic user commands into executable machine actions. Primarily three main components are necessary to process user inputs in this regard:

**Large Language Model**: LLMs have advanced rapidly in recent years, driven by novel DL architectures, and the increasing availability of computational resources and large-scale training data. With billions of parameters, these models are pretrained on an extensive text corpora and demonstrate impressive performance across a wide range of natural language tasks [9,10]. For interaction through keyboard and text in natural language, an LLM alone is sufficient. However, for voice-based interaction additional components are required: a speech-to-text (STT) module to transcribe spoken input, and a text-to-speech (TTS) module to convert text into audio output. Alternatively, speech-to-speech (S2S) agents enable direct conversational interaction.

**Agent**: Standalone LLMs are not capable of interacting with external systems or utilizing sources beyond their pretraining data [11]. However, their functionality can be extended through the integration of tools. This extension transforms them from purely text-generative models into agents able of taking actions and making decisions [12,13]. As a result, agents may additionally take over supervisory tasks from operators. A standardized approach for giving LLMs access to external systems, data sources, and software is the open-source *Model Context Protocol* (MCP) [14]. MCP specifies a unified protocol that enables LLMs to connect seamlessly with external systems.

**IT/OT – Communication interface:** PLCs are real-time computers used in industrial automation. They read sensor inputs, perform computations, and control outputs such as motors or valves and are running with standardized programming paradigms defined in IEC 61131-3 [15]. Given the diverse data structures and formats originating from heterogeneous applications, devices, machines, and software, the OPC UA standard was established to improve interoperability and industrial data exchange [5]. Through OPC data blocks, an OPC client can read, write, or monitor real-time data from an OPC server, that runs on a PLC. Whereas prior research has primarily focused on automated OPC UA information modeling [16] and PLC code [17–19] or test case [20] generation with LLMs, this paper investigates how the OPC UA communication standard can be leveraged to enable direct interaction with PLCs via natural language user commands.

## 3. Concept and experimental setup

In order to read or write machine variables semantically and through natural language, an agent with the ability to build an OPC UA connection to a machine's PLC is proposed. Machines in smart factories are mainly controlled through gaze, voice, tactile, gesture or haptic (touch) interaction [2]. However, in a first step the goal is to evaluate the tool-calling accuracy and their design. Consequently, STT and TTS components are not considered in the demonstrator setup and the experiments focus on the tool-calling capabilities of LLMs. Recent studies demonstrate that speech-enabled LLMs achieve remarkable performance in STT and TTS tasks [21]. Consequently, the evaluation of transcription and audio synthesis performance is not part of this research. Furthermore, kinesics interaction is not a reasonable modality for executing read or write operations in a factory environment. The following section describes the overall approach, the experimental demonstrator setup, and the evaluation methodology.

### *3.1. Conceptual approach*

Figure 2 schematically illustrates the conceptual approach to translate semantic user inputs into actions that read or write OPC UA nodes. A LLM is instructed by it's system prompt, which defines it's general role and tasks, and provides the corresponding context. The context consists of a <nodes_dictionary> and a <machine_credential_list>. The dictionary includes basic machine information, such as the names of the accessible variables, their OPC NodeIds, and data types. The NodeId functions as a unique identifier to ensure a reference in an OPC UA address space. An exemplary node can be, for instance, a variable with the name “temperature”, NodeId “ns=4;i=12” and datatype “Int16”. The <machine_credential_list> includes the corresponding OPC server endpoints and potential safety credentials, which are necessary to build the connection to the OPC UA server.

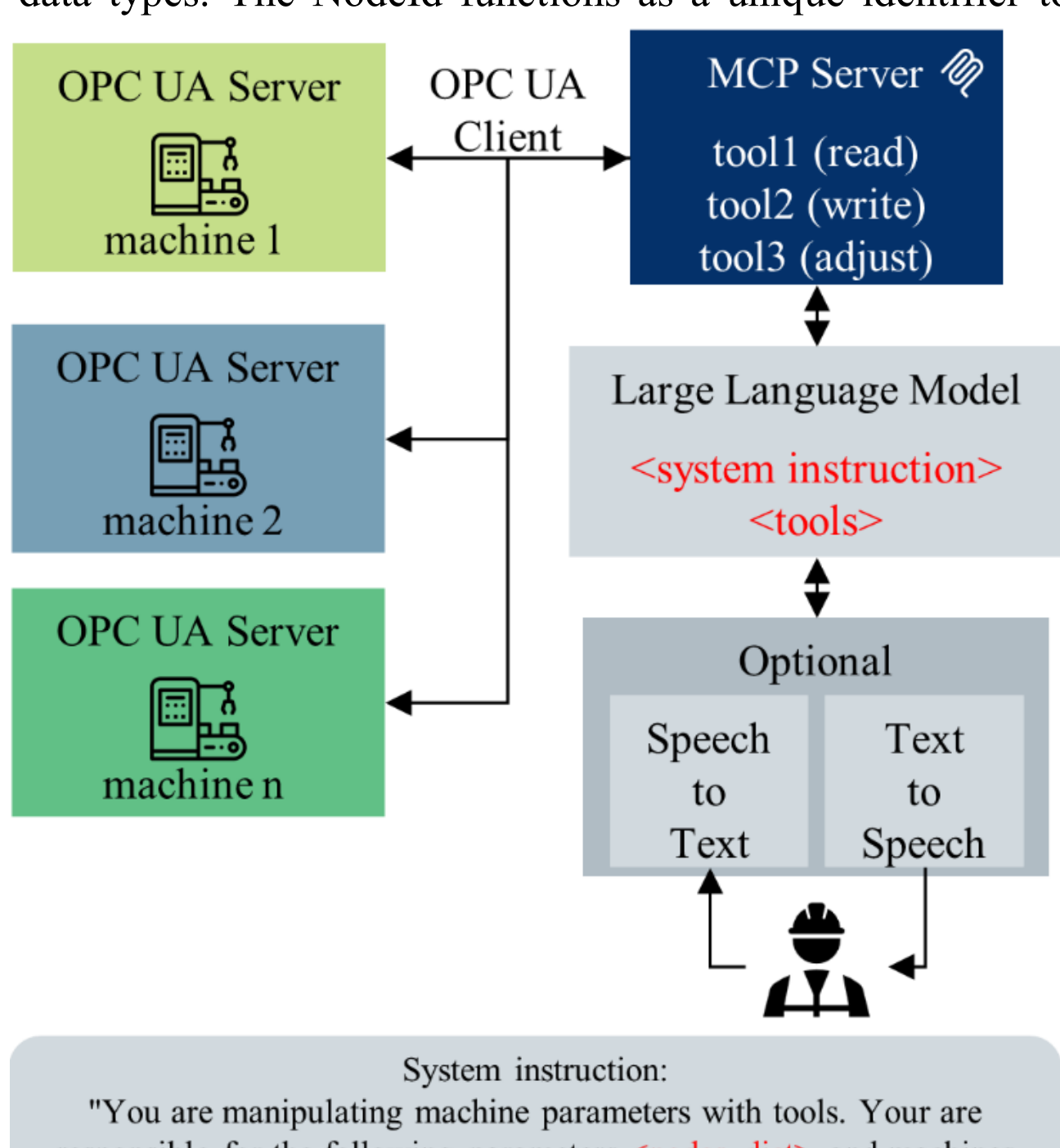

**Figure 2**. Conceptual architecture of the proposed approach for natural language machine control.

Moreover, the LLM is equipped with three distinct tools (read, write and adjust). These tools are python functions and can either be provided with “function calling” through a specific schema or by an MCP server. While the core logic for connecting to an OPC UA server, accessing its nodes, and performing read or write operations is encapsulated within the python functions, the LLM itself is responsible for correctly interpreting user input, selecting the appropriate function, and providing the correct arguments for the tool calls. The advantage of providing the tools in an MCP server lies in their central availability. By hosting them centrally, any LLM across different departments can be equipped with the ability to read and write OPC UA nodes, while the machine-specific information continues to be provided locally through the system instruction. The reading tool allows to retrieve the current state of a node, enabling requests such as “What is the current temperature?”. Additionally, the LLM can use this tool to verify whether a variable adjustment was successfully applied. The writing tool gives the model the functionality to overwrite the value of a variable, for example “Set temperature to 80 °C.” The adjustment tool extends the skillset of the model to execute relative changes in absolute or percentage values, such as “Reduce the temperature by 10 %.” While SOTA LLMs are capable of solving mathematical calculations, the combination of these three tools ensures that the LLM itself does not need to compute values. This reduces the risk of hallucinations, keeps the operation deterministic, enables the use of smaller models, and lowers computational resource requirements.

*3.2. Experimental setup of the demonstrator system*

To evaluate the conceptual approach, a demonstrator setup with a Siemens SIMATIC S7-1500 PLC was configured with four exemplary parameters:

- “motorspeed”, “ ns=4;i=11”, “Float”
- “temperature”, “ns=4;i=12”, “Int16”
- “textfield1“, “ns=4;i=14”, “String”
- “textfield2“, “ns=4;i=13”, “String”

A local workstation hosts the MCP server and runs the OPC UA and LLM client, as schematically shown in Figure 3. The MCP server was implemented with *Gradio*, and the workstation is directly connected to the Siemens PLC over TCP/IP. Three proprietary models (*GPT-5*, *GPT-5 mini*, *GPT-5 nano* with snapshot “2025-08-07” [22]) and two open-weight models (*GPT-oss:20b* [23], *Qwen3:32B* [24]) were utilized for the experimental setup. The latter were hosted locally with *Ollama* on a separate workstation with two *NVIDIA L40s* GPUs. While the proprietary models support MCP tool integration, the functionalities of the open-weight models were extended using the *OpenAI* function schema style.

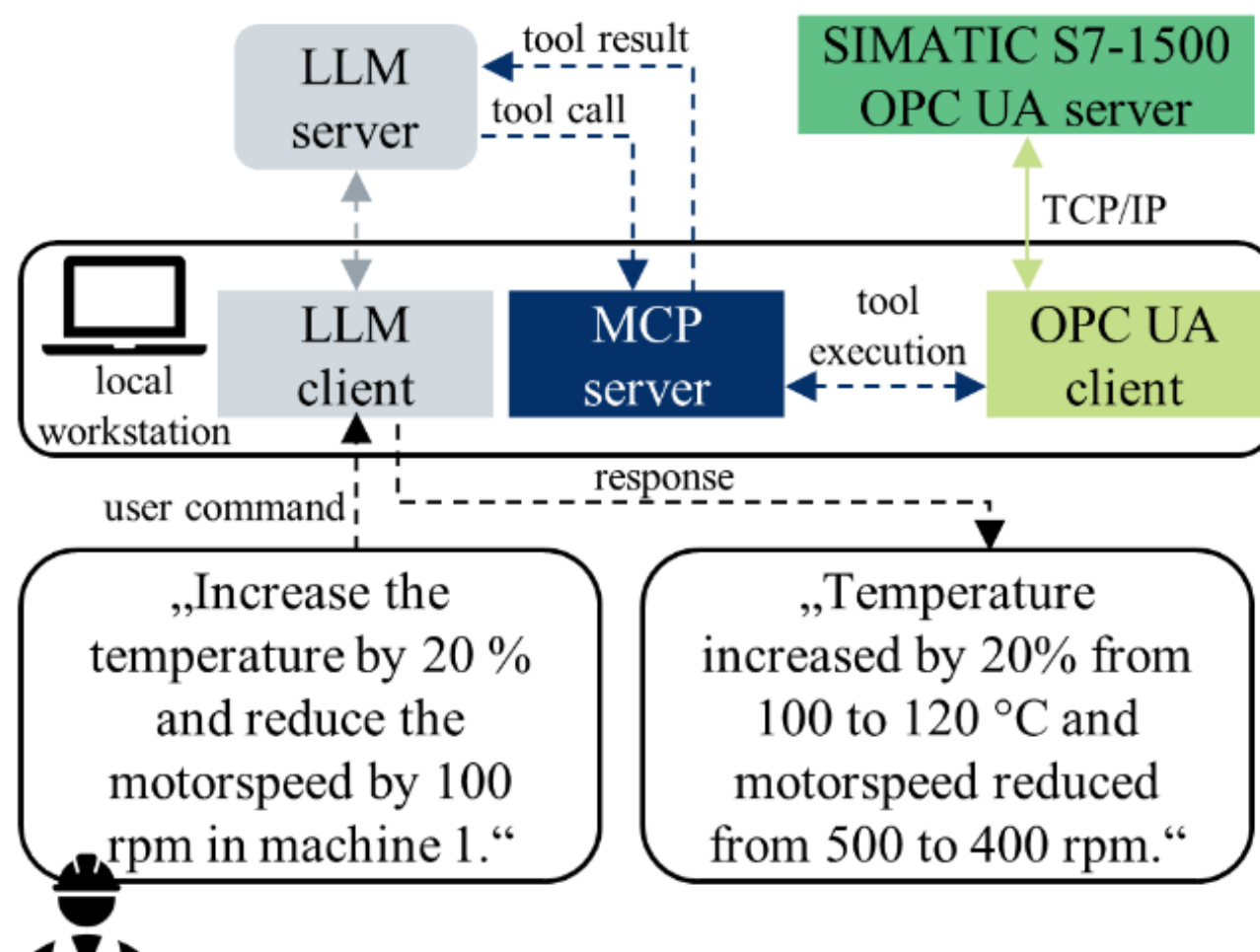

**Figure 3**. Schematic diagram of the experimental setup for evaluating the approach.

When a user issues a command to the LLM client on the local workstation, the prompt is forwarded to the LLM server (either local or cloud-based) for processing. Subsequently, the LLM determines the relevant tool and appropriate arguments. The tool execution is carried out either on the MCP server or locally through function calling, depending on the model’s capability. By executing the function, the OPC client connects to the PLC’s OPC server and performs the read, write or adjust operation. Afterwards, the tool result is returned, appended to the message history, and sent back to the LLM. Finally, the LLM processes the message again and returns either the final response to the user or chooses to call another tool.

To evaluate the actions and responses of the five models, 50 commands with four difficulty levels were generated with *GPT-5 Chat* and reviewed by the authors. The first difficulty level with 15 commands involved changing a single parameter, the second level again with 15 commands involved two parameters, followed by 10 commands involving three parameters, and another 10 commands with all four parameters. Exemplary commands are shown in Table 1. The state of all four parameters was logged after each prompt for evaluation. An action was considered correct if the resulting parameter states matched the expected ground truth. Accuracy was used as the evaluation metric, defined as the ratio of correct actions to the total number of commands. Furthermore, correctness was evaluated under a strict all-or-nothing criterion: an action was considered correct only when every affected parameter was set accurately. For instance, in the third command of Table 1, if *textfield1* and *textfield2* were updated correctly but *temperature* was not, the entire command was marked as incorrect. Since the commands were prompted as a sequence, follow-up errors can occur. For example, if *motorspeed* was set incorrectly in one command, a subsequent command to adjust the speed would also deviate from the ground truth. In such cases, the follow-up command was still counted as correct if the adjustment itself was executed properly, with only the ground truth being misaligned. Additionally, callback questions, instead of direct execution, were considered incorrect as well.

**Table 1.** Example commands of varying complexity, affecting different numbers of parameters simultaneously. Parameters include motorspeed (m), temperature (t), textfield 1 (tf1), and textfield 2 (tf2).

| **command** | **parameter** | | | |
|---|---|---|---|---|
| | **m** | **t** | **tf1** | **tf2** |
| 1: Raise motorspeed by 30 | x | | | |
| 2: Add +10 speed and +5°C | x | x | | |
| 3: tf1 = "Warning", tf2 = "Critical", temperature = 0 | | x | x | x |
| 4: Drop speed by 50%, lower temp by 100, tf1 = "Three", tf2 = "Four" | x | x | x | x |

## 4. Results and discussion of the demonstrator setup

In total, 50 commands were prompted to each of the five models, with the results summarized in Figure 4. The proprietary GPT models achieved accuracy scores of 96.0 % for *GPT-5* and *GPT-5 nano*, and 98.0 % for *GPT-5 mini*, corresponding to 48 respectively 49 out of 50 correct executions. In comparison, the open-weight models both achieved an accuracy of 90.0 %, or 45 out of 50 correct executions. The results show that the high-performance model *GPT-5* does not outperform its smaller counterparts *GPT-5 mini* and *GPT-5 nano*. Consequently, smaller sized proprietary models can be considered for tool-calling approaches like this. While the results of the open-weight models were competitive, they underperformed the proprietary GPT models. The findings underscore the importance of robust tool design, as misinterpretations persist even for models with large parameter counts. Five distinct error types were observed, categorized and summarized in Table 2.

Most of the errors (n = 9) were caused by incorrect execution of a tool, primarily due to omission of the minus operator. For example, in the 26th command ("Reduce speed by 10 and write 'Reset' in tf2") *GPT-5* did not reduce the *motorspeed* by 10, it increased it by 10, because the delta argument in the adjust tool was set to 10 instead of -10. Three errors occurred because the models executed tools twice, likely due to a misinterpretation of the appended tool result in the message history. For instance, in the 20th command ("Increase speed by 20 and set tf2 = 'Done'") *Qwen3:32b* executed both tools twice. This led to the *motorspeed* being increased by 40 instead of 20 and to two redundant adjustments of *textfield2*, which did not affect the final output. In one case, *GPT-5 mini* asked for confirmation of its interpretation for the 48th command. Two errors resulted from a misinterpretation of the operation verb. For instance, in 24th command ("Drop motorspeed 5 and tf1 = 'Empty'") *GPT-5 nano* misinterpreted the verb "drop" as "set". Consequently, it did not decrease the *motorspeed* by 5 but instead set the value to 5. Finally, one error appeared from misinterpreting a tool itself. In the 19th command ("Adjust motorspeed to 30 and reduce temperature by 100") *GPT-oss:20b* performed the setting operation with the *delta* argument of the adjust tool instead of using the *value* argument of the write tool. As a result, the *motorspeed* was increased by 30 instead of setting it to 30.

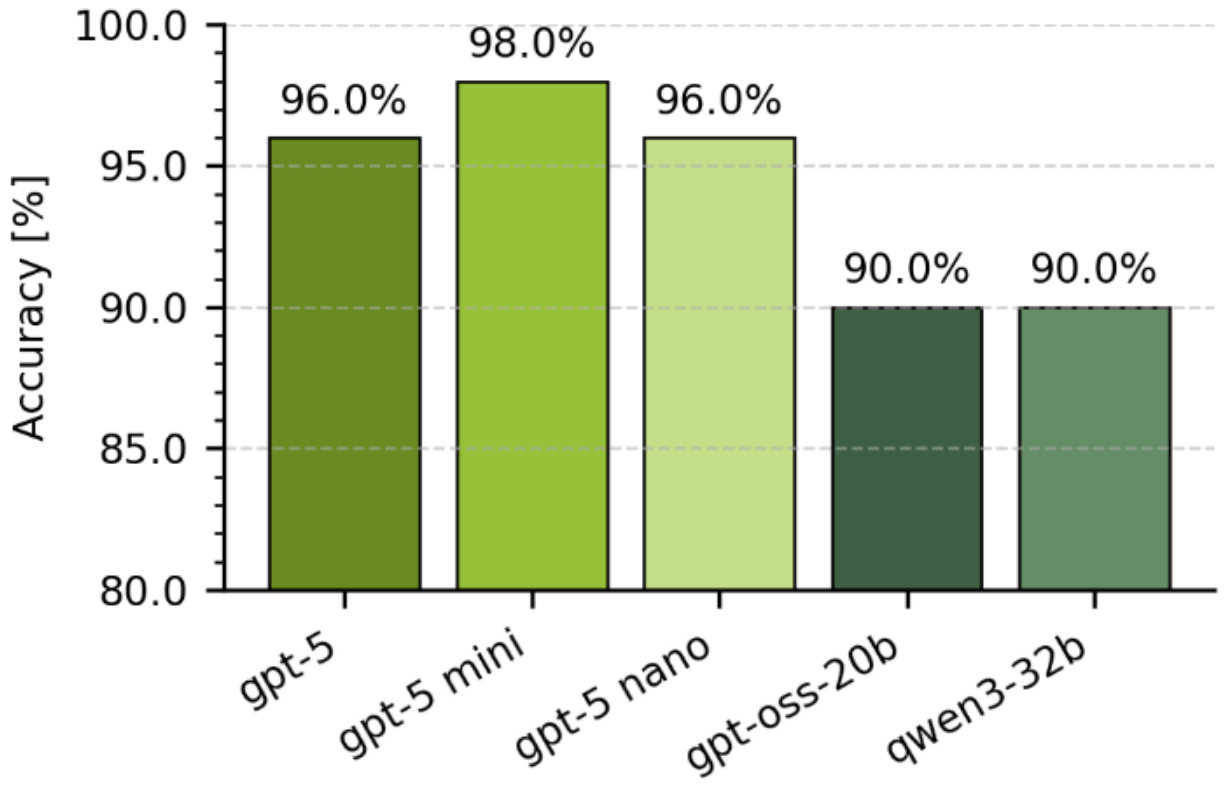

**Figure 4.** Accuracy comparison of different language models in executing machine control commands.

Overall, the tool design is with accuracies from 90 % to 98% is robust and the general approach is capable of controlling machines in natural language. Smaller and locally hosted models can be considered, especially when privacy concerns or latency requirements are of high interest. Nevertheless, safety levels or human approvals can additionally be integrated.

## 5. Transfer and integration into an industrial spray-coating machine

After testing the general concept, the approach is transferred to an industrial machine for spray-coating of wire bundles, according to [25]. The machine consists of a UR10e robot with application tool for polymers and a workpiece holder for a cable bundle. The polymer is applied as the robot moves the applicator over the section of the bundle to be coated. The deployed setup is shown in Figure 5.

The machine consists of various adjustable process parameters, among others: *coating length*, *application speed*, *nozzle distance*, *spraying pressure* or *material pressure*. Each parameter has four OPC UA sub-variables, including the maximum and minimum value, the current state and the input. Consequently, beside the writing capabilities, the model is additionally able to utilize it's reading function to retrieve more information about the current process state.

Moreover, an action variable was implemented that allows the model to trigger complete process sequences, such as *Wipe Nozzle* or *Apply Coating*. In addition, a feedback variable was provided through which the model can read status and failure messages from the PLC. For improved readability and international comprehensibility, the original German parameter names and the use cases below were translated into English throughout this paper. A schematic overview of the system architecture and process flow is illustrated in Figure 6.

**Table 2.** Categorization of error types observed during command execution across different models. Models include GPT5 (m1), GPT5-mini (m2), GPT5-nano (m3), GPT-oss:20b (m4), and Qwen3:32b (m5).

| Error type | m1 | m2 | m3 | m4 | m5 |
|---|---|---|---|---|---|
| | command index | | | | |
| 1: Incorrect tool execution due to sign error | 1,26 | | | 1,9, 37, 48 | 9,47, 48 |
| 2: Incorrect interpretation of tool feedback leading to repeated tool execution | 1 | | | | 20, 21 |
| 3: Uncertain interpretation requiring user confirmation | | 48 | | | |
| 4: Misinterpretation of operation verb (e.g., drop, increase, set, adjust) | | | 24, 48 | | |
| 5: Misinterpretation of tool | | | | 19 | |

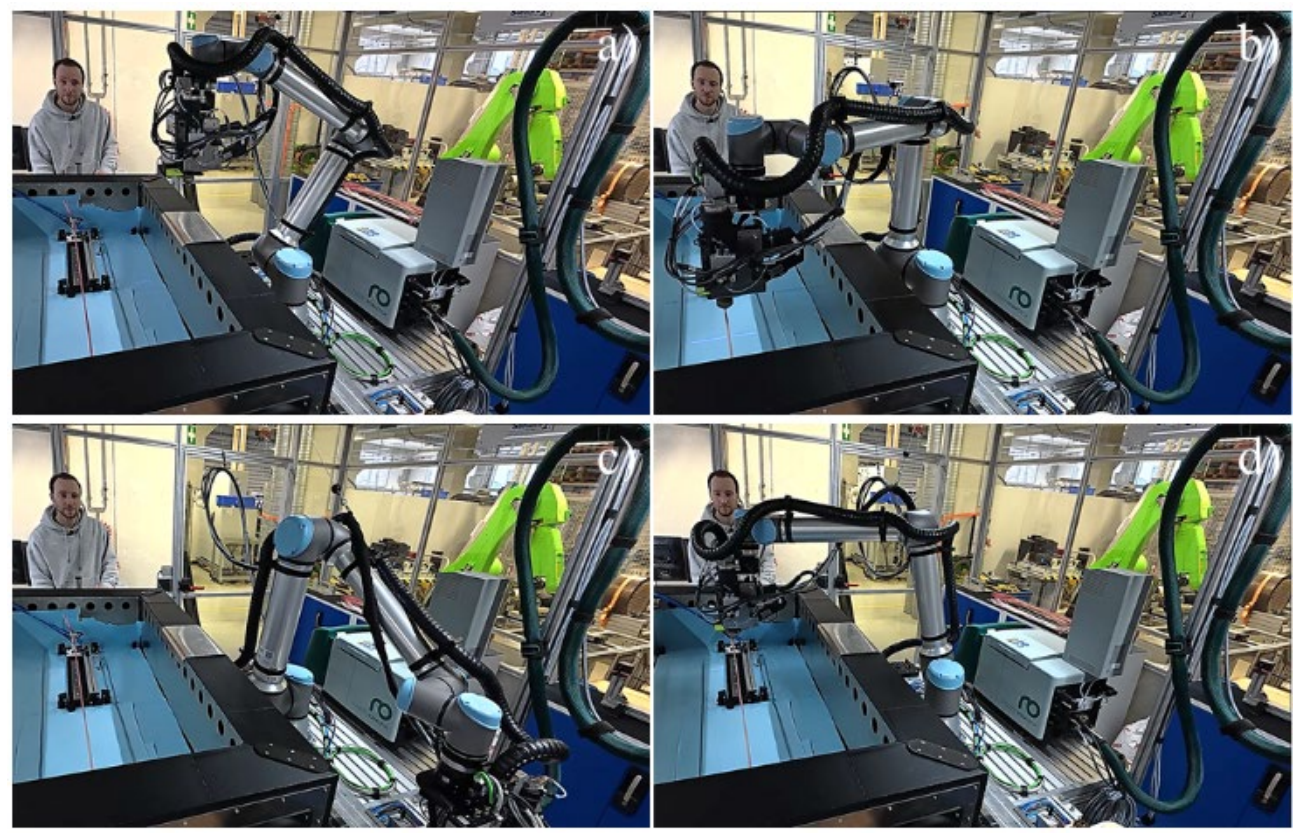

**Figure 5.** Deployed system at different stages of the interaction: a) initial state, b) after the verbal command to coat the wire, c) after the verbal command to clean the nozzle, and d) after the verbal command to coat the wire again with increased speed and length.

The setup is hosted locally with *llama.cpp* and one *NVIDIA L40s* GPU. *Whisper large-v3* [26] is utilized as STT-model and the TTS synthesis was performed using *Qwen3 TTS* [27]. As mentioned before user queries as well as parameter description are mostly conducted in German language. *Qwen3VL-8B-Instruct-F16* [28] functions as the language model and is utilized with an in-context learning approach. The model is adjusted for two specific downstream tasks by changing the content of its context window. Consequently, only one language model needs to be hosted, but it is queried two times per user input. First, it receives the user's request and operates as a tool-calling agent. Second, it processes the tool execution results and formulates the final answer, which, is subsequently synthesized by the TTS-model. The exemplary input of Figure 6 "Start the process, but with 20% higher speed." is recorded and saved as a .wav-file. Afterwards, it is transcribed in text by the STT-model. Together with the *system prompt (functions)* and the functions schema the transcribed text is forwarded to *Qwen3*. The model's output is the function call, which in this case results in:

- *adjust_variable*(<application speed>, <application_speed_nodeid>, <Int16>, <20>, <percent>)
- *write_variable*(<action>, <action_nodeid>, <string>, <Apply Coating>)

The output text is parsed and the two python functions are executed with the given arguments. The functions connect to the OPC UA server of the PLC as a client and adjust the parameters through their corresponding NodeIds. While each process parameter has it's own OPC UA variable, all actions only have one string variable in total. The python function writes, for instance, the string "Apply Coat" in the action variable, which the PLC is able to recognize as a defined process sequence. Afterwards, the PLC is sending the corresponding job number to the robot control and the robot is executing the task. Simultaneously, the tool result of the python functions, the original user input and the *system prompt (answer)* is processed by *Qwen3* to formulate the final answer. The output text is forwarded to the TTS model and a .wav-file is created. The file is carried out and the user receives "I increased the speed by 20% and started the process" as a feedback of the machine in natural language.

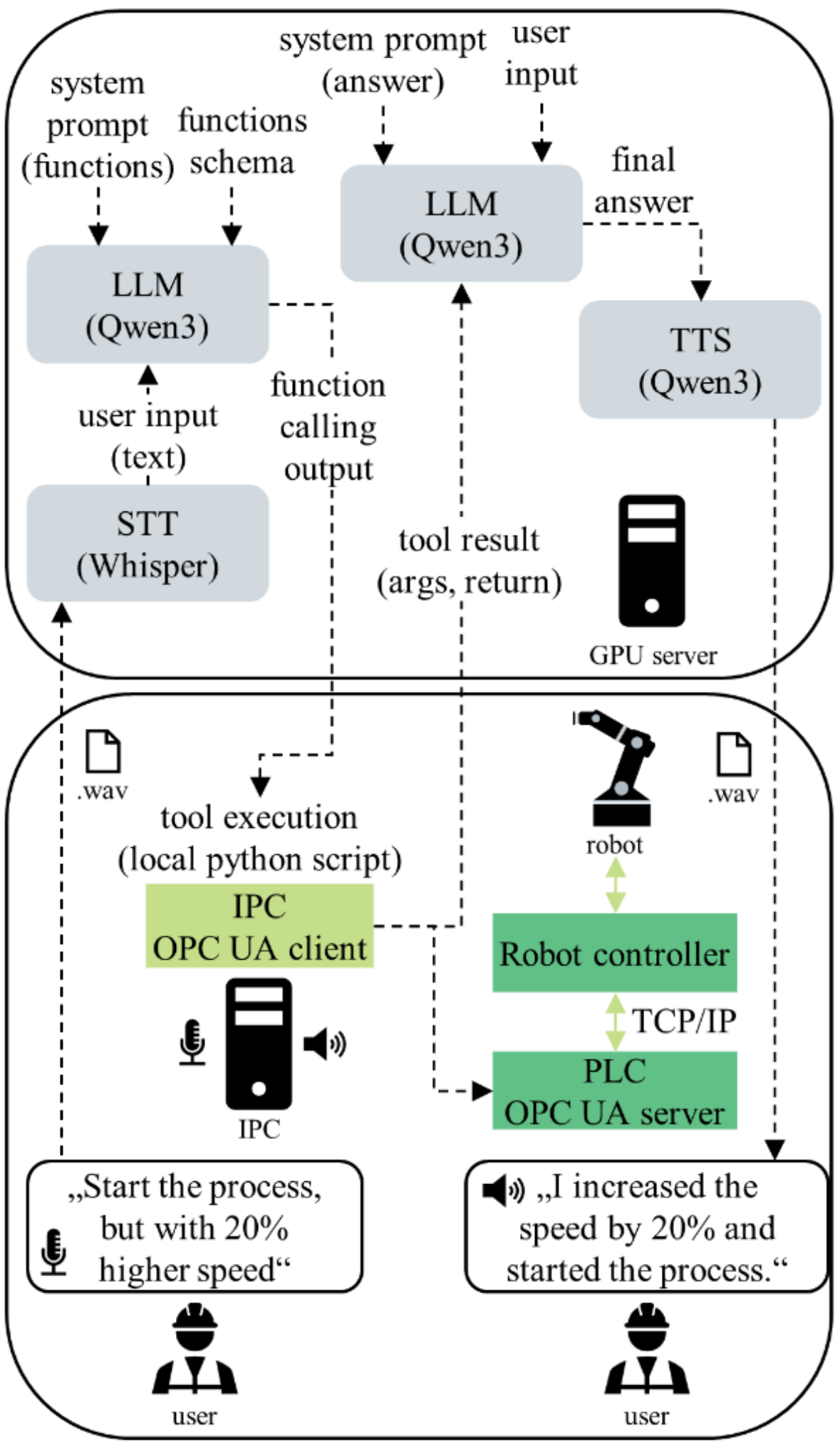

**Figure 6.** System architecture of the deployment in the spray coating machine.

## 6. Conclusion and outlook

This paper proposes an agent-based approach for modifying machine parameters through natural language, potentially replacing or supplementing SOTA touch interaction in industrial human-machine interfaces. The advantage lies in conversational interaction, which reduces the complexity of menu navigation and can accelerate operational workflows. The approach is universally designed and can therefore be applied to any machine that supports the OPC UA standard.

By equipping LLMs with tools, user commands can be translated into interactions with OPC UA nodes, allowing machine parameters to be read, written, or adjusted either absolutely or by percentage. This approach enables machine operators to interact with equipment more intuitively, and solely semantically trough natural language. The experimental results demonstrate that even smaller models are already capable of performing this task and calling functions correctly, achieving accuracies between 90.0 % and 98.0 %, with 45 to 49 of 50 commands executed correctly. Additionally, the concept has been transferred to real-world industrial machine for spray-coating to further validate the industrial implication. Future research should go beyond the interaction with humans and build complete closed-looped agentic systems for industrial production.

## Declaration of generative AI and AI-assisted technologies in the writing process

During the preparation of this work the authors used generative AI and AI-assisted technologies in order to improve the language and readability of their paper. After using this tool/service, the authors reviewed and edited the content as needed and take full responsibility for the content of the publication.

## Acknowledgement

The research presented in this paper is part of the project "KIdoka: AI-based quality agent for autonomous quality control through proactive fault detection and troubleshooting in cable production" (DIK-2408-0033// DIK0633/01), which is funded by the Bavarian Ministry of Economic Affairs, Regional Development and Energy (StMWi) and supervised by "VDI/VDE-IT". The authors would like to express their sincere gratitude to "Kühne + Vogel Prozessautomatisierung Antriebstechnik GmbH" for their support in preparing the experimental hardware setup.

## Data availability

The functions and schemas, the fifty generated commands and logged responses, as well as a demo video of the deployed approach in the spray-coating machine are available at [https://github.com/BJhof/llm_opcua_tools](https://github.com/BJhof/llm_opcua_tools).